\documentclass[preprintnumbers,amsmath,amssymb,aps,prd]{revtex4}
\usepackage{amsmath}
\usepackage[mathcal]{eucal}
\usepackage{latexsym}
\usepackage{amssymb}
\usepackage{color}
\newcommand*{\eps}{{\rlap{\lower2ex\hbox{$\,\,\tilde{}$}}{\epsilon_{ijk}}}}
\newcommand*{\EPS}{{\rlap{\lower2ex\hbox{$\,\,\tilde{}$}}{\epsilon_{i'j'k'}}}}
\newcommand*{\lmq}{{\rlap{\lower2ex\hbox{$\,\,\tilde{}$}}{\epsilon_{lmq}}}}
\newcommand*{\jmq}{{\rlap{\lower2ex\hbox{$\,\,\tilde{}$}}{\epsilon_{jmq}}}}
\newcommand*{\jql}{{\rlap{\lower2ex\hbox{$\,\,\tilde{}$}}{\epsilon_{jql}}}}
\newcommand*{\jlm}{{\rlap{\lower2ex\hbox{$\,\,\tilde{}$}}{\epsilon_{jlm}}}}
\newcommand*{\imq}{{\rlap{\lower2ex\hbox{$\,\,\tilde{}$}}{\epsilon_{imq}}}}
\newcommand*{\iql}{{\rlap{\lower2ex\hbox{$\,\,\tilde{}$}}{\epsilon_{iql}}}}
\newcommand*{\ilm}{{\rlap{\lower2ex\hbox{$\,\,\tilde{}$}}{\epsilon_{ilm}}}}
\newcommand*{\lmn}{{\rlap{\lower2ex\hbox{$\,\,\tilde{}$}}{\epsilon_{lmn}}}}
\newcommand*{\abc}{{\rlap{\lower2ex\hbox{$\,\,\tilde{}$}}{\epsilon_{abc}}}}
\newcommand*{\N}{{\rlap{\lower2ex\hbox{$\,\,\tilde{}$}}{N}}}
\newcommand{\tN}{{\rlap{\lower2ex\hbox{$\,\,\tilde{}$}}{N}}}
\newcommand*{\tM}{{\rlap{\lower2ex\hbox{$\,\,\tilde{}$}}{M}}}
\newcommand*{\imn}{{\rlap{\lower2ex\hbox{$\,\,\tilde{}$}}{\epsilon_{imn}}}}
\begin{document}
\title{Gravitational waves in intrinsic time geometrodynamics}

\author{Eyo Eyo Ita III}\email{ita@usna.edu}
\address{Physics Department, US Naval Academy. Annapolis, Maryland}
\author{Chopin Soo}\email{cpsoo@mail.ncku.edu.tw}
\address{Department of Physics, National Cheng Kung University, Taiwan}
\author{Hoi-Lai Yu}\email{hlyu@gate.sinica.edu.tw}
\address{Institute of Physics, Academia Sinica, Taiwan}

\bigskip

\begin{abstract}
Gravitational waves are investigated in Intrinsic Time Geometrodynamics. This theory has a non-vanishing physical Hamiltonian generating intrinsic time development in our expanding universe, and four-covariance is explicitly broken by higher spatial curvature terms. Linearization of Hamilton's equations about the de Sitter solution produces transverse traceless excitations, with the physics of gravitational waves in Einstein's General Relativity recovered in the low curvature low frequency limit. A noteworthy feature of this theory is that gravitational waves always carry positive energy {\it density}, even for compact spatial slicings without any energy contribution from boundary Hamiltonian.
This study of gravitational waves in compact $k= +1$ cosmological de Sitter spacetime is in contradistinction to, and complements, previous $k= -1$ investigations of Hawking, Hertog and Turok and other more familiar $k=0$ works. In addition, possible non-four-covariant Horava gravity contributions are considered (hence the use of canonical Hamiltonian, rather than Lagrangian, methods). Recent explicit $S^3$ transverse-traceless mode spectrum of Lindblom, Taylor and Zhang are also employed to complete the discussion.

%{Publication details:\\
%{\it New Commutation Relations for Quantum Gravity}, \\Chopin Soo and Hoi-Lai Yu, Chin. J. Phys. 53, 110106 (2015),\\
%Chinese Journal of Physics (Volume 53, Number 6 (November 2015))
%Special Issue On the occasion of 100 years since the birth of Einstein's General Relativity}\\
%Preprint: {\bf arXiv:1501.06282[gr-qc]}
%{\bf Keywords}:{\it Gravitational waves, Horava gravity theories, de Sitter perturbations, $S^3$ harmonics}
\end{abstract}

\maketitle

\section{Introduction}

Intrinsic Time Geometrodynamics (ITG) is a framework for geometrodynamics without the paradigm of space-time covariance which has been advocated in a series of works\cite{SOOYU,SOOYU1,SOOITAYU,CS}. A recent thorough discussion of the use of cosmic time and the resultant reduced phase space and effective Hamiltonian can be found in Ref.\cite{DOF}.
Equipped with spatial-diffeomorphism-invariant physical Hamiltonian, it resolves `the problem of time' and bridges  the deep divide between quantum mechanics and conventional canonical formulations of quantum gravity with a Schrodinger equation which describes first-order evolution in global intrinsic time. Einstein's theory of General Relativity which can be considered as a special case of a wider class of Horava gravity theories is recaptured at low curvatures and long wavelengths.

In Intrinsic Time Quantum Geometrodynamics (ITQG)\cite{SOOITAYU} the fundamental dynamical variables are the unimodular spatial 3-metric $\bar{q}_{ij}$ and the traceless momentric variable $\bar{\pi}^i_j$.  They are related to the standard General Relativity(GR) phase space variables, the three metric and
conjugate momentum $(q_{ij},\widetilde{\pi}^{ij})$ by
\begin{eqnarray}
\label{VARI}
\bar{q}_{ij}=q^{-1/3}q_{ij};~~\bar{\pi}^i_j=q^{1/3}\bar{q}_{jm}\bigl(\widetilde{\pi}^{im}-\frac{1}{3}q^{im}\widetilde{\pi}\bigr).
\end{eqnarray}
\noindent
$\bar{\pi}^i_j$ is the traceless part of the momentric variable (first introduced by Klauder\cite{Klauder}), and  the fundamental commutation relations for ITQG expressed through these variables are\cite{SOOITAYU,CSYU}
\begin{eqnarray}
\label{VARI1}
&[\bar{q}_{ij}(x),\bar{q}_{kl}(y)]=0,~~[\bar{q}_{ij}(x),\bar{\pi}^k_l(y)]=i\hbar \bar{E}^k_{lij}\delta(x-y),\cr
&[\bar{\pi}^i_j(x),\bar{\pi}^k_l(y)]=\frac{i\hbar}{2}\bigl(\delta^k_j\bar{\pi}^i_l-\delta^i_l\bar{\pi}^k_j\bigr)\delta(x-y);
\end{eqnarray}
\noindent
wherein $\bar{E}^i_{jmn}=\frac{1}{2}\bigl(\delta^i_m\bar{q}_{jn}+\delta^i_n\bar{q}_{jm}\bigr)-\frac{1}{3}\delta^i_j\bar{q}_{mn}$ is a traceless projector which also plays the role of the vielbein for the $(+,+,+,+,+)$ supermetric
${\bar G}_{mnpq} = \bar{E}^i_{jmn}\bar{E}^j_{ipq}$. It is noteworthy that the commutation relations of the momentric variables ${\bar\pi}^i_j$ are in fact the $su(3)$ algebra.
The physical Hamiltonian $H_{Phys}$ which generates evolution with respect to intrinsic time $T$ has been elucidated elsewhere\cite{ZPE,SOOITAYU}, and it takes the simple form
\begin{eqnarray}
\label{HAMIL}
H_{\rm Phys}=\frac{1}{\beta}\int  \bar{H}(x) d^3x,
\end{eqnarray}
\noindent
with a local Hamiltonian of density weight one\footnote{As shown in Ref.(3), the Hamiltonian density is the square root of a positive semi-definite, self-adjoint operator that governs the evolution of the theory with respect to intrinsic time $T = \frac{2}{3}\ln(V/V_0)$ wherein $V$ is the spatial volume of the universe. Einstein's Ricci scalar potential and cosmological constant term emerge after regularization of a coincident commutator term\cite{ZPE}.}
\begin{eqnarray}
\label{HAMIL1}
\bar{H}=\sqrt{\bar{\pi}^m_n\bar{\pi}^n_m+\alpha q(R-2\Lambda)+g^2\tilde{C}^{mn}\tilde{C}_{mn}} \,,
\end{eqnarray}
\noindent
and $\alpha = - \frac{1}{(2\kappa)^2}$.  The Cotton-York tensor density (of weight one) is denoted by $\tilde{C}^{mn}$ and $g$ (dimensionless) is a coupling constant of the theory.  It follows that $H_{Phys}$ is invariant under spatial diffeomorphisms, and the super-momentum constraint, $H_i =0$,  can be added to the total Hamiltonian of the theory. $H_{Phys}$ is not a local constraint, but a true non-vanishing Hamiltonian  which generates physical evolution of the variables $({\bar q}_{ij}, {\bar \pi}^i_j)$ with respect to the change $\delta T= \frac{2}{3}\delta\ln V$, wherein $V$ is the spatial volume of our universe\cite{SOOYU,SOOYU1,SOOITAYU,CS}. This is a very physical description of dynamics which resolves the `problem of time' in GR and its extensions, and renders them amenable to the usual rules of classical and quantum dynamics. Unlike many Horava gravity theories\cite{Horava} with an extra ambient time parameter, here $T$ is constructed from the intrinsic geometry of the 3-metric - a degree of freedom has been used: the determinant of the  metric, $q$, obeys the Heisenberg equation of motion $\frac{d\ln q^{1/3}}{dT}= 1$, and the trace of the momentum, ${\tilde {\pi}}^i_i$, is totally absent  in $H_{Phys}$. Thus, despite not having a local Hamiltonian constraint (as in `projectable' Horava gravity theories with an extra, and possibly pathological, mode), only two degrees of freedom ($({\bar q}_{ij}, {\bar \pi}^i_j)$ with $H_i =0$) are subject to fluctuations. An integrated Hamiltonian $H_{Phys}$ rather than a local Hamiltonian constraint also ensures that addition of higher spatial curvature terms does not lead to intractable second class constraints and/or inconsistencies in the constraint algebra.
Einstein's GR is the limit $\beta^2=1/6$ and $g=0$ i.e. when the potential term in ${\bar H}$ reduces to just the spatial Ricci scalar and cosmological constant terms. Without 4-covariance and arbitrary {\it a priori} lapse function $N$, Einstein's theory is recaptured in the sense that $H_{Phys}$ produces an effective or emergent lapse
while the EOM and constraints of GR lead  to precisely this same {\it a posteriori} value of the lapse function
\footnote{For any ADM decomposition of the metric, $N=\frac{{\sqrt q}(\partial_t \ln q^{1/3} -\frac{2}{3}\nabla_i N^i)}{4\beta\kappa {\bar H}}$ .This in fact ensures the Hamiltonian constraint is satisfied classically in the form $\frac{(Tr K)^2}{9} = \frac{4\kappa^2\beta^2}{q}{\bar H}^2$. Further details on the equivalence between GR and $H^{ADM}_{Phys}$ can be found in Refs.\cite{SOOYU,SOOYU1,CS}.}; and the square-root form of the Hamiltonian in (\ref{HAMIL1}) is needed for this agreement\cite{SOOYU,SOOYU1,CS}.
The presence of higher {\it spatial} curvature terms needed for improved UV convergence and completion of the theory signals the explicit loss of 4-covariance. A corresponding  Lagrangian of the Baierlein-Sharp-Wheeler type can be found\cite{SOOYU}, but it is rather cumbersome to work with, and not really needed. The Hamiltonian description of mechanics is both complete and consistent.
The use of canonical Hamiltonian, rather than Lagrangian methods, is more suitable for taking into account possible non-4-covariant contribution such as the Cotton-York term.  This also provides a sound canonical prescription for the study of gravitational waves in Horava gravity theories maintaining two gravitational degrees of freedom\cite{DOF}. A comparison of ITG with other approaches such as York extrinsic time and scalar field time can be found in the remarks section of Ref.\cite{DOF}.

\subsection{Hamilton's equations}

In this work classical gravitational wave equations will be derived through the Hamilton's equations. The Heisenberg equations of motion for the unimodular metric variable and momentric,
\begin{equation}
\frac{\partial\bar{q}_{ij}(x)}{\partial{T}}=\frac{1}{i\hbar}[\bar{q}_{ij}(x),H_{Phys}]; \qquad \frac{\partial\bar{\pi}^k_l(x)}{\partial{T}}=\frac{1}{i\hbar}[\bar{\pi}^k_l(x),H_{Phys}],
\end{equation}
lead\footnote{An explicit metric representation of the momentric operator is  $\bar{\pi}^i_j =\frac{\hbar}{i}{\bar E}^i_{jmn} \frac{\delta}{\delta \bar{q}_{mn}} $.}, in the classical limit of setting $\hbar \rightarrow  0$, to
\begin{equation}
\label{HAMIL3}
\frac{\partial\bar{q}_{ij}(x)}{\partial{T}}=\frac{1}{\beta\bar{H}(x)}\bar{E}^l_{kij}(x)\bar{\pi}^k_l(x),
\end{equation}
and
\begin{equation}
\label{HAMIL4}
\frac{\partial\bar{\pi}^k_l(x)}{\partial{T}}=-\frac{1}{\beta}\bar{E}^k_{lij}(x)\int \,d^3x'\frac{1}{\bar{H}(x')}\biggl[\frac{\alpha{q}(x')}{2}\frac{\delta{R}(x')}{\delta\bar{q}_{ij}(x)}
+\tilde{C}_{mn}\frac{\delta\tilde{C}^{mn}}{\delta\bar{q}_{ij}(x)}\biggr].
\end{equation}
\noindent
 In the above, we have used the fact that traceless part of the momentric commutes with the kinetic operator, $\bar{\pi}^m_n\bar{\pi}^n_m$  of $H_{Phys}$, which is a Casimir invariant of the $su(3)$ algebra generated by
$\bar{\pi}^i_j$.  Formulas collected in the Appendix yield
\begin{eqnarray}
\label{HAMIL5}
\frac{\partial\bar{\pi}^k_l(x)}{\partial{T}}
&=&-\frac{\alpha}{2\beta}q^{1/3}(x)\int \, d^3x'\frac{q(x')}{\bar{H}(x')}\bar{E}^k_{lij}(x)
\Bigl[-R^{ij}(x')+\nabla^i_{x'}\nabla^j_{x'}-q^{ij}(x')\nabla^2\Bigr]\delta(x-x')
\nonumber\\
&-&\frac{1}{\beta}\bar{E}^k_{lij}(x)\int \,d^3x'\frac{1}{\bar{H}(x')}
\tilde{C}_{mn}\frac{\delta\tilde{C}^{mn}}{\delta\bar{q}_{ij}(x)}.
\end{eqnarray}
\noindent
Besides projecting out the traceless part of the Ricci tensor, $\bar{R}^k_l=R^k_l-\frac{1}{3}\delta^k_lR$, the traceless projector also annihilates the $\nabla^2$ term.
Upon integration, the result is
\begin{equation}
\label{HAMIL6}
\frac{\partial\bar{\pi}^k_l(x)}{\partial{T}}
=\frac{\alpha{q}}{2\beta\bar{H}}\bar{R}^k_l-\frac{\alpha}{2\beta}q^{4/3}\bar{E}^k_{lij}\nabla^i\nabla^j\frac{1}{\bar{H}}
-\frac{1}{\beta}\bar{E}^k_{lij}(x)\int d^3x'\frac{1}{\bar{H}(x')}\tilde{C}_{mn}\frac{\delta\tilde{C}^{mn}}{\delta\bar{q}_{ij}(x)}.
\end{equation}

.
\section{Gravitational waves on de Sitter background}

\subsection{Background solution and linearization}
We shall consider background solutions of the Hamilton equations with constant spatial 3-curvature geometries compatible with the Cosmological Principle. Explicitly, these Robertson-Walker 3-metrics are
\begin{eqnarray}
d\ell^2=a^2(T)[\frac{dr^2}{1-kr^2}+r^2(d\theta^2+\sin^2\theta d\phi^2)]
\end{eqnarray}
\noindent
with $k=+1,0,-1$ respectively for the compact $S^3$, and non-compact $\mathbb{R}^3$ and $H^3$ intrinsic 3-geometries.
The metric is Einstein, $R_{ij}=\frac{1}{3}q_{ij}R$ with $R=\frac{6k}{a^2}$; hence $\bar{R}_{ij}$ and the Cotton-York tensor $\tilde{C}_{ij}$ both vanish. In addition, $\bar{q}_{ij}=q^{-1/3}q_{ij}$ is independent of $a$ (and thus of $T$).
For the background extrinsic geometry, we set the traceless momentric variable $\bar{\pi}^i_j$ to zero.  These considerations lead to spatially covariantly constant Hamiltonian density $\bar{H}=\sqrt{\alpha{q}(R-2\Lambda)}$, and each term in (\ref{HAMIL6}) vanishes, resulting in $\dot{\bar{\pi}}^i_j=0$. Thus, the pair of Hamilton equations (\ref{HAMIL3}) and (\ref{HAMIL4}) are identically satisfied, the initial data is preserved; and the Robertson-Walker 3-geometry with vanishing momentric indeed constitutes a background solution of the theory, even when $H_{Phys}$ contains higher curvature terms $\tilde{C}^{mn}\tilde{C}_{mn}$ in addition to the scalar potential of Einstein's GR. The constant spatial curvature solution with vanishing momentric is also a saddle point of the exact vacuum solution in the Cotton-York era\cite{SOOITAYU}.

It is noteworthy that ${\bar H}$ involves only the square of $\tilde{C}^{mn}$  (which is identically zero for any constant 3-curvature metric). Thus we may state a simple theorem:  any spatially conformally flat solution of GR is also a solution of the EOM of the Hamiltonian of Eq.(\ref{HAMIL1})\cite{HCL}.

In the Arnowitt-Deser-Misner(ADM) decomposition of any 4-dimensional classical solution with coordinate time variable $t$, the lapse function takes the form $N=\frac{\sqrt{q} \partial_t \ln q^{1/3}}{4\beta\kappa \bar H}$ modulo spatial diffeomorphisms\cite{SOOYU,CS} .
We can therefore recast the background solution into the usual 4-dimensional Robertson-Walker form by reparametrizing the cosmic time interval as $dt^{\prime }:=Ndt =\frac{(\partial_t \ln a^2 )dt}{\sqrt{6}\beta\sqrt{\frac{\Lambda}{3}-\frac{k}{a^2}}}$. To wit,
\begin{eqnarray}
\label{3CUR}
ds^2=-dt^{\prime 2}+a^2(t^{\prime})[\frac{dr^2}{1-kr^2}+r^2(d\theta^2+\sin^2\theta d\phi^2)],
\end{eqnarray}
\noindent
and the above relation between $a$ and $t'$  reduces to $ \frac{da}{dt' } ={\sqrt 6}\beta\sqrt{\frac{\Lambda}{3} a^2 -k}$. This yields, for the GR value of $\beta =\sqrt\frac{1}{6}$, the de Sitter solution with $a(t') = \sqrt{\frac{3}{\Lambda}}\cosh[\sqrt{\frac{\Lambda}{3}}(t'- t'_0)], A e^{\sqrt{\frac{\Lambda}{3}}(t'-t'_0)} , \sqrt{\frac{3}{\Lambda}}\sinh[\sqrt{\frac{\Lambda}{3}} (t'- t'_0)]$  respectively for $ k =+1,0, -1$ spatial slicings.

\par
\indent
We shall linearize the Hamilton equations about the de Sitter background $(^{*}{\bar q}_{ij}, ^{*}{\bar\pi}^i_j)$ with $S^3$ slicings, and
expand the variables as ${\bar q}_{ij} =^{*}{\bar q}_{ij} +{\bar h}_{ij}$  and ${\bar\pi}^i_j = ^{*}{\bar\pi}^i_j + \Delta{\bar\pi}^i_j$ .
In addition, on account of spatial diffeomorphism symmetry, we require the physical  fluctuations to be transverse i.e. $^{*}\nabla^i {\bar h}_{ij}=0$; and ${\bar h}_{ij}$, being perturbations of the unimodular ${\bar q}_{ij}$, are traceless ($^*q^{ij}{\bar h}_{ij}=0$)  as well.

Differentiating with respect to $T$, the Euler-Lagrange equation for the metric fluctuation,
\begin{equation}
\label{RECAP3}
\frac{\partial^2\bar{q}_{ij}}{\partial{T}^2}=\frac{\partial}{\partial{T}}\Bigl(\frac{1}{\beta\bar{H}}\bar{E}^l_{kij}\Bigr)\bar{\pi}^k_l
+\frac{1}{\beta\bar{H}}\bar{E}^l_{kij}\frac{\partial\bar{\pi}^k_l}{\partial{T}},
\end{equation}
\noindent
yields the linearized identity
\begin{eqnarray}
\label{RECAP4}
\frac{\partial^2\bar{h}_{ij}}{\partial{T}^2}
=^*[\frac{\partial}{\partial{T}}\Bigl(\frac{1}{\beta\bar{H}}\bar{E}^l_{k(ij)}\Bigr)]{\Delta}\bar{\pi}^k_l
+\Delta\Bigl(\frac{\partial}{\partial{T}}\Bigl(\frac{1}{\beta\bar{H}}\bar{E}^l_{kij}\Bigr)\Bigr){^{*}\bar{\pi}^k_l}\nonumber\\
+^*\Delta\Bigl(\frac{1}{\beta\bar{H}}\bar{E}^l_{kij}\Bigr){^{*}\Bigl(\frac{\partial\bar{\pi}^k_l}{\partial{T}}\Bigr)}
+^{*}\Bigl(\frac{1}{\beta\bar{H}}\bar{E}^l_{kij}\Bigr)\frac{\partial(\Delta\bar{\pi}^k_l)}{\partial{T}}.
\end{eqnarray}
On account of vanishing $^{*}\bar{\pi}^i_j$, only the first and last terms remain. Linearization of (\ref{HAMIL3}) yields
\begin{equation}
\label{metric-momentric}
({\beta^*\bar{H}})\frac{\partial\bar{h}_{ij}(x)}{\partial{T}}=^*\bar{E}^l_{kij}\Delta\bar{\pi}^k_l,
\end{equation}
and substituting for $^*\bar{E}^l_{kij}\Delta\bar{\pi}^k_l$ in the first term of (\ref{RECAP4}) leads to the EOM for $\bar{h}_{ij}$  which is
\begin{equation}
\label{REC}
\frac{\partial^2\bar{h}_{ij}}{\partial{T}^2}   = -\frac{\partial \ln{^*{\bar H}}}{\partial T}(\frac{\partial{\bar h}_{ij}}{\partial T}) +
^{*}\Bigl(\frac{1}{\beta\bar{H}}\bar{E}^l_{kij}\Bigr)\frac{\partial(\Delta\bar{\pi}^k_l)}{\partial{T}}.
\end{equation}

\subsection{Gravitational wave equation}

Explicit calculations lead to
 $\frac{\partial \ln{^*{\bar H}}}{\partial T} = \frac{(R-3\Lambda)}{(R-2\Lambda)}$,
and
\begin{eqnarray}
\label{WAVE}
\frac{\partial(\Delta\bar{\pi}^k_l)}{\partial{T}}=-\frac{\alpha{q}}{4\beta^*\bar{H}}q^{km}\bigl(\nabla^2-\frac{1}{3}R\bigr){{h}_{ml}}+\dots \,.
\end{eqnarray}
\noindent
wherein by $\dots$ we mean the higher curvature contribution arising from $\tilde{C}^{mn}\tilde{C}_{mn}$. This shall be addressed later.
 The background Hamiltonian density $(\ref{HAMIL1})$ with zero momentric,  $^*\bar{H}$, is covariantly constant, and  $\frac{\alpha{q}}{^*\bar{H}^2}=\frac{1}{R-2\Lambda}$ with $R= \frac{6k}{a^2}$.
 Henceforth we drop the $*$ label when there is no confusion, and it is understood that, apart from the perturbations $({\bar h}_{ij}, {\bar\pi}^i_j)$, all other metric entities refer to the de Sitter background.

 With the use of the above identities, (\ref{REC}) gives the resultant gravitational wave equation on de Sitter background, and expressed with respect to intrinsic time $T$, as
\begin{equation}
\label{WAVE2}
\frac{\partial^2 \bar{h}_{ij}}{\partial{T}^2} +\frac{(R-3\Lambda)}{(R-2\Lambda)}\frac{\partial{\bar h}_{ij}}{\partial T}  +\frac{1}{4\beta^2(R-2\Lambda)}\bigl(\nabla^2-\frac{R}{3}\bigr)\bar{h}_{ij}+\dots=0
\end{equation}
for physical, transverse traceless perturbations ${\bar h}_{ij}$. Without the higher order curvature terms, this reproduces the gravitational wave equation for GR on a de Sitter background for $\beta^2 = 1/6$.
The factor $(R- 2\Lambda) = 6(\frac{k}{a^2} - \frac{\Lambda}{3})$ vanishes only at the de Sitter ``throat" at $t'=t'_0$ for $k=+1$, but this factor is otherwise always negative regardless of whether $k=+1,0,-1$.
Likewise,  the coefficient $\frac{(R-3\Lambda)}{(R-2\Lambda)}$ is positive definite, and the $\frac{\partial{\bar h}_{ij}}{\partial T}$ term plays the role of frictional force, tempered by the expansion of the universe.
Bearing in mind $\frac{da}{dt' } ={\sqrt 6}\beta\sqrt{\frac{\Lambda}{3} a^2 -k}$ , with $R=\frac{6k}{a^2}$, and $dT = 2d\ln a(t')$, conversion of Eq.(\ref{WAVE2}) into variation w.r.t. $t' $ yields
$\Big(\frac{1}{24\beta^2(\frac{\Lambda}{3} - \frac{k}{a^2})}\frac{\partial^2}{\partial t'^2}-\frac{1}{24\beta^2(\frac{\Lambda}{3} - \frac{k}{a^2})}\nabla^2\Big) {\bar h}_{ij}  + ... =0$ which implies the speed of the GR wave is 1 (in units c =1 since we have previously used the notation $-dt'^2$ instead of $-c^2dt'^2$ in the de Sitter metric).

\subsection{Higher curvature contribution}

The higher curvature contribution to the wave equation which arises from (\ref{HAMIL6}) is
\begin{equation}
\label{NONLINEAR}
-\frac{1}{\beta^2}\frac{1}{^*{\bar H}(x)}{^{*}}(\bar{E}^l_{kij}\bar{E}^k_{luv})(x)\int d^3x'\frac{1}{^{*}\bar{H}(x')}
(\Delta \tilde{C}_{mn}(x')^{*}\frac{\delta\tilde{C}^{mn}(x')}{\delta \bar {q}_{uv}(x)}),
\end{equation}
and it can be explicitly computed as in the Appendix. This may in turn be expressed as
\begin{equation}
\label{NONLINEAR2}
-\int d^3x'\frac{q^{-\frac{1}{3}}}{\beta \bar{H}(x')}\Delta \tilde{C}_{mn}(x')\mathcal{O}^{mn}\,_{ij}\delta(x-x'),
\end{equation}
\noindent
with $\mathcal{O}^{ijkl}
:=-\frac{1}{\beta \bar H}[\frac{g}{8}(q^{ik}\epsilon^{jlm}+q^{jk}\epsilon^{ilm}+q^{il}\epsilon^{jkm}
+q^{jl}\epsilon^{ikm})\nabla_m](\nabla^2-\frac{R}{3})$ for the background.
Upon integration, the equation with higher curvature contribution is,
\begin{equation}
\label{NONLINEAR1}
\frac{\partial^2 \bar{h}_{ij}}{\partial{T}^2}+\frac{(R-3\Lambda)}{(R-2\Lambda)}\frac{\partial{\bar h}_{ij}}{\partial T}+\frac{1}{4\beta^2(R-2\Lambda)}(\nabla^2-\frac{R}{3})\bar{h}_{ij}+{\mathcal{O}^{\dagger}}^{mn}\,_{ij} \mathcal{O}_{ mn}\,^{kl}\bar h_{kl}=0,
\end{equation}
\noindent
wherein explicit  the higher curvature term is ${\mathcal{O}^{\dagger}}^{mn}\,_{ij} \mathcal{O}_{ mn}\,^{kl}{\bar h}_{kl} = \frac{q} {\beta^2 {\bar H}^2}\Big(-\frac{g^2}{4}(\nabla^2-\frac{R}{2})\Big)(\nabla^2-\frac{R}{3})^2{\bar h}_{ij}
$.

At this level of approximation, we may assume $dT =\frac{2}{3}d\ln V= 2d\ln a =-d\ln R$, or $e^{T-T_{now}}= (\frac{a}{a_{now}})^2 = (1+z)^{-2} = \frac{R_{now}}{R}$. Thus the equation may be expressed entirely in terms of $T$, $a$ or $z$-development.
While all $k=0,\pm1$ are valid descriptions in this work\cite{Turok}  in which $dT =2d\ln a$, in general ITQG uses $dT=\frac{2}{3}d\ln V$,  thus favoring compact manifolds with finite spatial volumes.

In fact ${\bar h}_{ij}$ can be explicitly expanded in terms of $k=+1$ or $S^3$ tensor (density)\footnote{The eigenvalues of the Laplacian operator are not affected by multiplication with any power of $q$ which is covariantly constant.}
  harmonics of Ref.\cite{Lindblom}, $\bar Y^{Klm}_{(4,5)ij}$, with $K \geq 2$ . These are the two orthogonal transverse traceless eigenfunctions of Laplacian operator, $\nabla^2$, with (negative) eigenvalues $E'_K = \frac{2-K(K+2)}{a^2} = \frac{R[2-K(K+2)]}{6}$. A similar expansion can be done for the transverse traceless  $\Delta\bar{\pi}^a_b$ .  Through ${\bar h}_{ab} = \sum_{I=4,5} C^{(I)}_{Klm}\bar Y^{Klm}_{(I)ab}$ and orthogonality of the eigenfunctions, (\ref{NONLINEAR1}) reduces to an equation for intrinsic time-dependence of the mode coefficients which carry discrete eigenvalues $\{Klm\}$. The resultant equation which encodes full-fledged information of all time dependence of the physical modes arising from gravitational perturbations during different epochs of the expanding de Sitter universe is
\begin{equation}
\label{NONLINEAR2}
\ddot{ C}_{\{K\}}+\frac{(R-3\Lambda)}{(R-2\Lambda)}\dot{ C}_{\{K\}}+\frac{E_K}{4\beta^2(R-2\Lambda)}{ C}_{\{K\}}- \frac{ {g^2}(E_K+ \frac{R}{6})}{4\beta^2\alpha (R-2\Lambda)}E_K^2{ C}_{\{K\}}=0;
\end{equation}
\noindent
wherein, for simplicity, we have denoted $C_{\{K\}}:=C^{(I)}_{Klm}$; and defined $E_K := E'_K - \frac{R}{3} = -\frac{K(K+2)R}{6}$ which is the eigenvalue of $\nabla^2 -\frac{R}{3}$.

\section{Energy of gravitational perturbations, and further remarks}

ITQG, as in Horava-type gravity theories, introduces only higher order spatial, but not time, derivatives into the wave equation through higher order spatial curvature terms which improve the ultra-violet convergence of the theory without compromising unitarity; yet the theory captures the physics of Einstein's GR in long wavelength low curvature circumstances. From the wave equation, we can also see that the propagator for flat background will contain additional terms (up to the highest order of ${1/p^6}$ from the square of Cotton-York tensor in $H_{Phys}$), but there will be no additional poles for $p_0$ in the absence of higher time derivatives.  In the modified wave equation, the ratio of the higher curvature contribution to that of Einstein's GR is explicitly
$-\frac{[ {g^2}(E_K +\frac{R}{6})]E_K}{\alpha} = (16\pi)^2{[{g^2}(K+1)^2]K(K+2)}(\frac{l_{\rm Planck}}{a})^4$, which can be computed for any given set of $K,R, b$ and $g$. In the current epoch, this ratio too small to be of significance at LIGO's characteristic detection wavelengths. However, departures from Einstein's theory can become significant in the regime of large curvatures in the early universe and/or for large values of $K$. In the era of ${a}\rightarrow 0$, all physics is dominated by the Cotton-York term, the de Sitter solution is a saddle point of the exact vacuum state\cite{SOOITAYU}; and, instead of Einstein's GR, Eq.(\ref{NONLINEAR2}) will be dominated by the last term associated with linearized excitations in the Cotton-York era.

  There is no contradiction in having both {\it physical local energy density} and {\it spatial} diffeomorphism invariance. Without the paradigm of 4-covariance,
  the total Hamiltonian density is not required to vanish; at each point, 2 of the d.o.f. remain even after spatial diffeomorphisms are taken into account. In perturbative excitations, the remaining physical d.o.f. are precisely the transverse traceless $({\bar h}_{ij}, \Delta{\bar\pi}^i_j)$ modes.
A noteworthy feature of ITQG in which time change is identified with variation in the logarithm of  (finite) spatial volume is that gravitational waves always carry physical positive energy {\it density}, even for compact spatial slicings without any energy contribution from boundary Hamiltonian. The energy for the gravitational wave excitation is
  \begin{eqnarray}
 &&H_{\rm Phys}[{\bar\pi}^i_j, q_{ij}] - H_{\rm Phys}[^*{\bar\pi}^i_j, ^*q_{ij}]\cr \approx \int \frac{1}{ {2\beta}{^*}\bar H}&\Big[&\Delta\bar{\pi}^m_n\Delta\bar{\pi}^n_m
    + \frac{\alpha q} {4}{\bar q}^{ik}{\bar q}^{jl}{\bar h}_{ij}(\nabla^2 - \frac{R}{3}){\bar h}_{kl}\cr &&+(\beta{\bar H})^2(q^{\frac{1}{3}}\mathcal{O}_{mn}\,^{kl}{\bar h}_{kl})(q^{\frac{1}{3}}\mathcal{O}^{mnij}{\bar h}_{ij})\Big]d^3x.
 \end{eqnarray}
The expression is positive-definite for $k=0$ and $+1$  since $\alpha < 0$, but the $R =6k/a^2$  term is negative for $k=-1$. Again by expanding in eigenmodes of the Laplacian operator, the expression of the Hamiltonian density can be computed explicitly in terms of the mode coefficients. In the classical theory, the momentric is related to the time change of the metric via Eq.(\ref{metric-momentric}).
  %The energy density of the excitations obeys the dispersion relation,  ${\bar H} =\sqrt{\Delta\bar{\pi}^m_n\Delta\bar{\pi}^n_m + \alpha ^*q(*R-2\Lambda)}$.

While it is true that adopting a d.o.f. as `clock' can yield a non-vanishing local Hamiltonian even when 4-covariance is maintained, multi-fingered  `time' suffers from ordering problems\cite{Wheeler}  and clock-dependent alternative histories.
ITQG, or in this regard, Horava gravity theories, are not gauge-fixed versions of Einstein's GR; they have true Hamiltonians, global time evolutions and `preferred slicings'.  Positive-definite spatial metric  bequeaths space-like
separation, ``a notion of `simultaneity' and a common moment of a rudimentary `time'"\cite{Wheeler}. Dynamical fields evolve, expansion of our universe is a `time' change, and energy associated with that generator $H_{\rm Phys}$ is physical. This study of gravitational waves in compact $k= +1$ cosmological de Sitter spacetime is in contradistinction to, and complements, previous $k= -1$ investigations of Hawking, Hertog and Turok\cite{Turok} and other more familiar $k=0$ works. In addition, possible non-four-covariant Horava gravity\cite{Horava} contributions are considered (hence the use of canonical Hamiltonian, rather than Lagrangian, methods).  To complete the discussion,
recent explicit $S^3$ transverse-traceless mode spectrum of Lindblom, Taylor and Zhang\cite{Lindblom} are employed.

%``The universe allows no, and needs no, external clock"\cite{Smolin}, it and only it is all-encompassing and robust enough to be the universal and ultimate clock.

\section{Acknowledgments}
This work has been supported in part by the U.S. Naval
Academy, the Ministry of Science and Technology (R. O.
C.) under Grants No. MOST 105-2112-M-006-010 and
No. 106-2112-M-006-009, and the Institute of Physics,
Academia Sinica. E. E. I. would also like to acknowledge
the support of the University of South Africa (UNISA),
Department of Mathematical Sciences, and to express his
gratitude for the hospitality under the Visiting Researcher
Program.
\section{Appendix}

In three dimensions the Weyl curvature is zero, and the Riemann curvature
tensor $R^n_{mjk}=\partial_j\Gamma^n_{mk}-\partial_k\Gamma^n_{mj}+\Gamma^n_{js}\Gamma^s_{mk}-\Gamma^n_{ks}\Gamma^s_{mj}$ can be written completely in terms of the Ricci curvature through
\begin{equation}
\label{CURVATURE}
R^n_{mjk}=q_{mk}R^n_j+\delta^n_jR_{mk}-q_{mj}R^n_k-\delta^n_kR_{mj}+\frac{R}{2}\bigl(\delta^n_kq_{mj}-\delta^n_jq_{mk}\bigr).
\end{equation}
\noindent
Variation of the connection is given by
$
\delta\Gamma^m_{in}=\frac{1}{2}q^{mr}\Bigl(\nabla_i\delta{q}_{rn}+\nabla_n\delta{q}_{ri}-\nabla_r\delta{q}_{in}\Bigr),
$
while variation of the Ricci tensor and the curvature scalar result in
\begin{eqnarray}
\label{RELATION2}
\delta{R}_{ij}=\frac{1}{2}\Bigl(\nabla^n\nabla_i\delta{q}_{nj}+\nabla^n\nabla_j\delta{q}_{in}-\nabla^2\delta{q}_{ij}-q^{kl}\nabla_i\nabla_j\delta{q}_{kl}\Bigr);\nonumber\\
\delta{R}=\delta(q^{ij}R_{ij})=q^{ij}\delta{R}_{ij}-R^{ij}\delta{q}_{ij}=-R^{ij}\delta{q}_{ij}+\nabla^i\nabla^j\delta{q}_{ij}-q^{ij}\nabla^2\delta{q}_{ij}.
\end{eqnarray}
\noindent
Another useful result from commuting covariant derivatives involving a covariant divergence is
\begin{eqnarray}
\label{COVARIANT}
&\nabla^n\nabla_i\delta{q}_{nj}=\nabla_i\nabla^n\delta{q}_{nj}+R^l_i\delta{q}_{lj}-q^{mn}R^l_{jmi}\delta{q}_{nl} =
\nabla_i\nabla^n\delta{q}_{nj}-\frac{R}{2}\delta{q}_{ij}\cr &-q_{ij}R^{mn}\delta{q}_{mn}
+\bigl(2R^n_i\delta{q}_{nj}+R^n_j\delta{q}_{ni}\bigr)
-\bigl(R_{ij}-\frac{1}{2}q_{ij}R\bigr)\delta\hbox{ln}q,
\end{eqnarray}
\noindent
where we have used (\ref{CURVATURE}).  Putting these facts together, we have, for transverse-traceless variations with respect to the $S^3$ background that
\begin{equation}
\label{RELATIVE}
(\delta{R}_{ij})_{S^3}=\frac{1}{2}\Bigl(\frac{R}{2}\delta{q}_{ij}+\frac{R}{2}\delta{q}_{ij}-\nabla^2\delta{q}_{ij}-\nabla_i\nabla_j(^*q^{kl})\delta{q}_{kl}\Bigr)
=-\frac{1}{2}(\nabla^2-R)\delta{q}_{ij},
\end{equation}
and $(\nabla^n\nabla_i\delta q_{nj})_{S^3} =\frac{1}{2}R\delta q_{ij}$.

The Cotton-York tensor density of weight one $\tilde{C}^{ij}$ is the functional derivative of the Chern-Simons functional
\begin{equation}
\label{CURVATURE1}
W_{CS}=\frac{1}{4}\int \epsilon^{ikj}\Bigl(\Gamma^m_{in}\partial_j\Gamma^n_{km}+\frac{2}{3}\Gamma^m_{in}\Gamma^n_{js}\Gamma^s_{km}\Bigr)d^3x,
\end{equation}
i.e.
\begin{equation}
\frac{\delta W_{CS}}{\delta q_{ij}} =\tilde{C}^{ij}= \epsilon^{imn}\nabla_m (R^j_n -\frac{1}{4}R\delta^j_n) =
\frac{1}{2}(\epsilon^{imn}\nabla_m R^{j}_n +\epsilon^{jmn}\nabla_mR^{i}_n).
\end{equation}
That $\tilde{C}^{ij}$ is symmetric and the last equality above can be established through the Bianchi identity $\nabla_m (R^m_n -\frac{1}{2}R\delta^m_n) =0$. The corresponding functional variation of $\tilde{C}^{ij}$ is
\begin{eqnarray}
\label{CURVATURE6}
2\delta\tilde{C}^{ij}&=&(\epsilon^{imn}\nabla_m\delta{R}^j_n+\epsilon^{imn}\bigl((\delta\Gamma^j_{ms})R^s_n-(\delta\Gamma^s_{mn})R^j_s\bigr)+i\leftrightarrow{j})\nonumber\\
&=&\epsilon^{imn}\nabla_m\delta{R}^j_n+\frac{1}{2}\epsilon^{imn}\Bigl(q^{jr}\bigl(\nabla_m\delta{q}_{rs}+\nabla_s\delta{q}_{rm}-\nabla_r\delta{q}_{ms}\bigr)R^s_n\cr
&&-q^{sr}\bigl(\nabla_m\delta{q}_{rn}+\nabla_n\delta{q}_{rm}-\nabla_r\delta{q}_{mn}\bigr)R^j_s\Bigr)
+ i\leftrightarrow{j}.
\end{eqnarray}
\noindent
Taking into account the symmetrization of the indices $i,j$; and  $k,l$; we arrive at
\begin{equation}
\label{CURVATURE8}
\frac{\delta\tilde{C}^{ij}(x')}{\delta{q}_{kl}(x)}=\frac{1}{8}\Bigl(q^{ik}\epsilon^{jlm}+q^{jk}\epsilon^{ilm}+q^{il}\epsilon^{jkm}
+q^{jl}\epsilon^{ikm}\Bigr)\nabla_m\Bigl(\nabla^2-\frac{R}{3}\Bigr)\delta(x-x').
\end{equation}
\noindent
Thus,
\begin{eqnarray}
\label{CURVATURE9}
&\frac{1}{\beta \bar H}\frac{\delta\mathcal{C}^{ij}(x)}{\delta\bar{q}_{kl}(x')}=q^{\frac{1}{3}}\mathcal{O}^{ijkl}\delta(x-x')\cr
&=-\frac{q^{\frac{1}{3}}}{\beta \bar H}\Bigl[\frac{g}{8}\Bigl(q^{ik}\epsilon^{jlm}+q^{jk}\epsilon^{ilm}+q^{il}\epsilon^{jkm}
+q^{jl}\epsilon^{ikm}\Bigr)\nabla_m\Bigr](\nabla^2-\frac{R}{3})\delta(x-x');
\end{eqnarray}
\noindent
and ${\Delta\mathcal{C}^{ij}}(x) =\int \frac{\delta\tilde{C}^{ij}(x)}{\delta\bar{q}_{kl}(x')}{\bar h}_{kl}(x') d^3x'$. By explicit computations, it  follows that on $S^3$,
\begin{equation}
\label{CURVATURE10}
{\mathcal{O}^{\dagger}}^{mn}\,_{ij} \mathcal{O}_{ mn}\,^{kl}{\bar h}_{kl} = \frac{q}  {\beta^2 {\bar H}^2}\Big(-\frac{g^2}{4}(\nabla^2-\frac{R}{2})\Big)(\nabla^2-\frac{R}{3})^2{\bar h}_{ij}.
\end{equation}
%We also include for reference the expression for variation of $\bar H$ which is
%\begin{eqnarray}
%\label{POTENTIAL1}
%\Delta\bar{H}=\frac{1}{\bar{H}}\Big({^*\bar{\pi}^m_n}\Delta\bar{\pi}^n_m
%+\frac{\alpha}{2}q\Delta{R}+\frac{\alpha}{2}(R-2\Lambda)\Delta{q}
%+{^{*}\mathcal{C}^{mn}}\Delta\mathcal{C}_{mn}\Bigr).
%\end{eqnarray}
%\noindent

\end{document}